\documentclass[prl,aps,twocolumn,letterpaper,10pt,final,nofootinbib]{revtex4}
\usepackage{amsfonts}
\usepackage{amsmath}
\usepackage{amssymb}
\usepackage{graphicx}

\providecommand{\U}[1]{\protect\rule{.1in}{.1in}}

\input{tcilatex}
\begin{document}

\title{From qubits and actions to the Pauli-Schr\"{o}dinger equation }
\author{Salomon S. Mizrahi}
\affiliation{Departamento de F\'{\i}sica, CCET, Universidade Federal de S\~{a}o Carlos,
Caixa Postal 676, S\~{a}o Carlos, 13565-905, S\~{a}o Paulo,\textit{\ }Brazil}
\email{salomon@df.ufscar.br}
\date{\today}

\begin{abstract}
Here I show that a classical or quantum bit state plus one simple operation,
an \emph{action}, are sufficient ingredients to derive a quantum dynamical
equation that rules the sequential changes of the state. Then, by assuming
that a freely moving massive particle is the qubit carrier, it is found that
both, the particle position in physical space and the qubit state, change in
time according to the Pauli-Schr\"{o}dinger equation. So, this approach
suggests the following conjecture: because it carries one qubit of
information the particle motion has its description enslaved by the very
existence of the internal degree of freedom. It is compelled to be no more
described classically but by a wavefunction. I also briefly discuss the
Dirac equation in terms of qubits. \newline

Published in \textbf{Physica Scripta T 135 014007 (2009)}, \newline
doi:10.1088/0031-8949/2009/T135/014007
\end{abstract}

\maketitle

\section{Introduction}

Quantum mechanics (QM) can be instructed either by adopting the schemes
proposed by its inventors (Born, Heisenberg, Schr\"{o}dinger, Jordan) or,
more rigorously, following Dirac, within the Hilbert space framework, or
even using Feynman's path integral approach. As it is believed to be
pedagogically more appealing, almost all textbooks prefer to begin with the
nonrelativistic approach, discussing the wave-particle dualism,
wavefunctions, Schr\"{o}dinger equation, Hilbert space, non-commutative
operators, etc. However, looking at the emblematic dynamical equations of
Schr\"{o}dinger and Dirac, one notes that the Schr\"{o}dinger equation (SE)
is less fundamental than Dirac's relativistic equation, because this one
contains inherently the internal degree of freedom spin, while it is absent
in the former. In between there is the Pauli-Schr\"{o}dinger equation (PSE),
which was derived by Pauli when he applied the low energy approximation in
Dirac equation. Although being nonrelativistic yet the PSE is more complete
than SE because the spin is inherently present, while in the SE the spin
must be added as an extra degree of freedom. See references in \cite%
{HESTENES} for a detailed discussion.

In the beginning of the 1980's the possibility of quantum computation was
foreseen by people like Benioff, Feynman and Deutsch \cite%
{BENIOFF,FEYNMAN,DEUTSCH}, and their work influenced a mini-revolution that
began in the 1990\'s, which threw a new look in QM, mainly in its
interpretation and potentialities to explain new phenomena; in the last 15
years we have witnessed huge theoretical developments along with ingenious
experiments involving single or few atoms or molecules, electrons and
photons. So, the understanding of quantum physics has widen, shaping the new
arena called \emph{quantum information theory (QIT) }that borrowed many
concepts of classical information theory. In this context it seems to exist
a recognition \cite{FUCHS} that QM is a special kind of information theory
immersed in Hilbert space, and characterized by a reversible logic \cite%
{BENNETT,DIVICENZO,FREDKIN}.

In that connection, by using elementary concepts of communication theory, as
bits and gates, however represented in the framework of Hilbert space, I
will show here that a quantum evolution equation for one classical bit (%
\emph{Cbit}$,$\emph{\ }as defined in \cite{MERMIN}) or quantum bit (\emph{%
qubit}) of information can be derived. Then, by asking what could be the 
\emph{carrier} of one qubit (or spin\ 1/2), the natural choice is a particle
of mass \emph{m} characterized by its kinetic energy, and this information
is introduced in the qubit dynamical equation. This procedure is sufficient
to derive the PSE that rules the time evolution of both, the qubit and the
particle, its carrier. So, it ceases to be a particle in the Newtonian sense
to become a hybrid compelled to display wave properties and described by a
wavefunction. The qubit/spin evolution acquires an ascendancy over the
particle motion, being at the root of the observed quantum properties of
matter. Last but not least, the Dirac equation and its solution are briefly
discussed in terms of qubits.

\section{Cbits and actions}

In classical information theory the numbers in $\mathbb{Z}_{2}=\left\{
0,1\right\} $ are associated to bits, as in a relay or in a memory storage
device. One can go one step further and associate a particular
representation to the numbers 0 and 1: a column matrix for one classical bit
of information, the Cbit state, $1\longrightarrow \binom{1}{0}$ and $%
0\longrightarrow \binom{0}{1}$, as like the states \textquotedblleft
up\textquotedblright\ and \textquotedblleft down\textquotedblright\ for the
spin 1/2. These states can be written in the more familiar form of Dirac's
kets $\left\vert x\right\rangle $, $\left\vert \bar{x}\right\rangle $ ($\bar{%
x}=1-x$) for $\left\{ x,\bar{x}\right\} \in \mathbb{Z}_{2}$, and the bras
are the transposed, so $\mathcal{H}_{2}^{\times }$ $\equiv \left(
\left\langle 1\right\vert ,\left\langle 0\right\vert \right) $ is the dual
space of $\mathcal{H}_{2}\equiv \left( \left\vert 0\right\rangle ,\left\vert
1\right\rangle \right) $. The qubit $\binom{a}{b}$ is a generalization of
the Cbit, with $a$ and $b$ being complex numbers. The simplest operators to
be used are the identity $\boldsymbol{I}$, $\boldsymbol{I}\left\vert
x\right\rangle =\left\vert x\right\rangle $, and the NOT $\boldsymbol{X}$
that inverts the Cbit state, $\boldsymbol{X}\left\vert x\right\rangle
=\left\vert \bar{x}\right\rangle $. So the 4-uple $\mathcal{Q}=\left\{ 
\mathbb{Z}_{2},\mathcal{H}_{2},\mathcal{H}_{2}^{\times },\mathcal{L}%
_{2}\right\} $, $\mathcal{L}_{2}=\left\{ \boldsymbol{I},\boldsymbol{X}%
\right\} $ plus the field of complex numbers $\mathbb{C}$ are sufficient
tools for my purposes.

The \emph{action} $\boldsymbol{U}\left( \alpha ,\beta \right) \equiv \alpha 
\boldsymbol{I}+\beta \boldsymbol{X}$ is a linear map of a Cbit or qubit into
a qubit, $\boldsymbol{U}\left( \alpha ,\beta \right) \left\vert
x_{0}\right\rangle \rightarrow \left\vert x_{1}\right\rangle =\alpha
\left\vert x_{0}\right\rangle +\beta \left\vert \bar{x}_{0}\right\rangle $,
for arbitrary parameter $\alpha $, $\beta $ in $\mathbb{C}$. Let's first
restrict the parameters values to two numbers: $\beta =$ $\bar{\alpha}%
=1-\alpha $ and $\left\{ \alpha ,\bar{\alpha}\right\} \in \mathbb{Z}_{2}$,
such that $\boldsymbol{U}_{\alpha }\left\vert x_{0}\right\rangle \equiv
\alpha \left\vert x_{0}\right\rangle +\bar{\alpha}\left\vert \bar{x}%
_{0}\right\rangle $ is still a Cbit because $\alpha \bar{\alpha}=0$. The
actions $\left\{ \boldsymbol{U}_{0},\boldsymbol{U}_{1}\right\} $ form a
group: (a) the unit element is $\boldsymbol{I=U}_{0}$ $,$ while $\boldsymbol{%
X=U}_{1}$; (b) the inverse is $\boldsymbol{U}_{\alpha }^{-1}=$ $\boldsymbol{U%
}_{\alpha }$; (c) the product of two elements is an element in the group $%
\boldsymbol{U}_{\alpha _{2}}\boldsymbol{U}_{\alpha _{1}}=\boldsymbol{U}%
_{\beta }=\beta \boldsymbol{I}+\bar{\beta}\boldsymbol{X}$ with $\beta
=\alpha _{2}\alpha _{1}+\bar{\alpha}_{2}\bar{\alpha}_{1}$, and $\bar{\beta}=%
\overline{\alpha _{2}\alpha _{1}+\bar{\alpha}_{2}\bar{\alpha}_{1}}=\alpha
_{2}\bar{\alpha}_{1}+\bar{\alpha}_{2}\alpha _{1}$; (d) the associative
property $\left( \boldsymbol{U}_{\alpha _{3}}\boldsymbol{U}_{\alpha
_{2}}\right) \boldsymbol{U}_{\alpha _{1}}=\boldsymbol{U}_{\alpha _{3}}\left( 
\boldsymbol{U}_{\alpha _{2}}\boldsymbol{U}_{\alpha _{1}}\right) $ holds, and
the elements are unitary $\boldsymbol{U}_{\alpha }^{\dagger }=\boldsymbol{U}%
_{\alpha }^{-1}$.

Sequential $n$ actions 
\begin{equation}
\boldsymbol{U}_{n}\left( \vec{\alpha}\right) \equiv \prod_{j=1}^{n}\left(
\alpha _{j}\boldsymbol{I}+\bar{\alpha}_{j}\boldsymbol{X}\right) ,
\label{evol}
\end{equation}%
applied on a Cbit maps it into another Cbit, $\boldsymbol{U}_{n}\left( \vec{%
\alpha}\right) \left\vert x_{0}\right\rangle =\left\vert x_{n}\right\rangle $%
, going through the intermediary states $\left\{ \left\vert
x_{1}\right\rangle ,\left\vert x_{2}\right\rangle ,\left\vert
x_{3}\right\rangle ,...,\left\vert x_{n-1}\right\rangle \right\} $. Each set
of numbers $h_{n}=\left\{ \alpha _{n},...,\alpha _{3},\alpha _{2},\alpha
_{1}\right\} $ defines one \emph{history}, or \emph{trajectory}. One can
also write $\boldsymbol{U}\left( \alpha _{1}\right) \left\vert
x_{0}\right\rangle $ by changing the label of the Cbit state, and by a
trivial formal manipulation it is simple to show that $\boldsymbol{U}%
_{\alpha _{1}}\left\vert x_{0}\right\rangle =\left\vert x_{1}\right\rangle
=\left\vert \alpha _{1}x_{0}+\bar{\alpha}_{1}\bar{x}_{0}\right\rangle $,
identifying the label as the mapped bit $x_{1}\equiv \alpha _{1}x_{0}+\bar{%
\alpha}_{1}\bar{x}_{0}$. By induction $x_{n}=\alpha _{n}x_{n-1}+\bar{\alpha}%
_{n}\bar{x}_{n-1}$ for $n=1,2,3,...$. Also holds the transitivity property
expressed by the composition law 
\begin{equation}
\boldsymbol{U}_{n}\left( \vec{\alpha}^{\left( 2\right) }\right) \boldsymbol{U%
}_{m}\left( \vec{\alpha}^{\left( 1\right) }\right) =\boldsymbol{U}%
_{n+m}\left( \vec{\alpha}\right) .  \label{trans1}
\end{equation}%
The sequence of actions (\ref{evol}) is reversible since each one is
unitary, then $\left\vert x_{0}\right\rangle =\boldsymbol{U}_{n}^{-1}\left( 
\vec{\alpha}\right) \left\vert x_{n}\right\rangle $. The reverse history is
given by the sequence $h_{n}^{-1}=\left\{ \alpha _{1},...,\alpha
_{n-2},\alpha _{n-1},\alpha _{n}\right\} $. In summary, (\ref{evol}) carries
the evolution $\left\vert x_{0}\right\rangle \longrightarrow $ $\left\vert
x_{n}\right\rangle $, and $\boldsymbol{U}_{n}^{-1}\left( \vec{\alpha}\right)
=\prod_{j=n}^{1}\left( \alpha _{j}\boldsymbol{I}+\bar{\alpha}_{j}\boldsymbol{%
X}\right) $ does the inverse path, $\left\vert x_{n}\right\rangle
\longrightarrow \left\vert x_{0}\right\rangle $. Formally, $\boldsymbol{U}%
_{n}\left( \vec{\alpha}\right) $ and $\boldsymbol{U}_{n}^{-1}\left( \vec{%
\alpha}\right) $ are the same since each factor in Eq. (\ref{evol}) commutes
with all others.

\subsection{ Coefficients on a circle of unit radius}

I now assume the parameters $\alpha $ and $\beta $ in $\boldsymbol{U}\left(
\alpha ,\beta \right) $ to be real, with $\alpha \beta \neq 0$ and ask $%
\alpha ^{2}+\beta ^{2}=1$, so $\left( \alpha ,\beta \right) \in \mathbb{%
\tilde{R}}_{2}$ is the set of all real numbers on a circle of radius $1$. As
so, acting on a Cbit one gets a qubit, $\boldsymbol{U}\left( \alpha ,\beta
\right) \left\vert x_{0}\right\rangle =\alpha \left\vert x_{0}\right\rangle
+\beta \left\vert \bar{x}_{0}\right\rangle $. Two consecutive operations
give $\boldsymbol{U}\left( \alpha _{2},\beta _{2}\right) \boldsymbol{U}%
\left( \alpha _{1},\beta _{1}\right) =\boldsymbol{U}\left( \alpha _{3},\beta
_{3}\right) $, and as $\alpha _{1}^{2}+\beta _{1}^{2}=$ $\alpha
_{2}^{2}+\beta _{2}^{2}=1$, it follows that $\alpha _{3}^{2}+\beta
_{3}^{2}=1+4\alpha _{2}\alpha _{1}\beta _{2}\beta _{1}\neq 1$, so $\left(
\alpha _{3},\beta _{3}\right) \notin \mathbb{\tilde{R}}_{2}$ and $%
\boldsymbol{U}\left( \alpha _{3},\beta _{3}\right) $ is \emph{not} an
element of the group, unless one of the four coefficients is zero, therefore
any probabilistic interpretation for $\alpha ^{2}$ and $\beta ^{2}$ fails.
Moreover, the inverse action is $\boldsymbol{U}^{-1}\left( \alpha ,\beta
\right) =\tilde{\alpha}\boldsymbol{I}+\tilde{\beta}\boldsymbol{X}$, where
the new parameters $\tilde{\alpha}=\alpha /\left( \alpha ^{2}-\beta
^{2}\right) $ and $\tilde{\beta}=-\beta \left( \alpha ^{2}-\beta ^{2}\right) 
$ are \emph{not} in $\mathbb{\tilde{R}}_{2}$. Due to the reality of $\alpha $
and $\beta $, $\boldsymbol{U}\left( \alpha ,\beta \right) $ is a
self-adjoint operator $\boldsymbol{U}^{\dagger }\left( \alpha ,\beta \right)
=\boldsymbol{U}\left( \alpha ,\beta \right) $ however it is\textit{\ }not
unitary since $\boldsymbol{U}^{\dagger }\left( \alpha ,\beta \right) \neq 
\boldsymbol{U}^{-1}\left( \alpha ,\beta \right) $. Although the norm $%
\left\Vert \boldsymbol{U}\left( \alpha ,\beta \right) \left\vert
x_{0}\right\rangle \right\Vert =1$ is parameter independent, this is not
true for the inverse $\left\Vert \boldsymbol{U}^{-1}\left( \alpha ,\beta
\right) \left\vert x_{0}\right\rangle \right\Vert =\left\vert \alpha
^{2}-\beta ^{2}\right\vert ^{-1}$. Thus, if we want to construct an
evolution operator $\boldsymbol{U}_{n}\left( \vec{\alpha},\vec{\beta}\right)
=\prod_{j=1}^{n}\left( \alpha _{j}\boldsymbol{I}+\beta _{j}\boldsymbol{X}%
\right) $, with $\alpha _{j}^{2}+\beta _{j}^{2}=1$, that is also reversible,
we are in trouble. Since the inverse of $\boldsymbol{U}\left( \alpha
_{j},\beta _{j}\right) $ is $\boldsymbol{U}\left( \tilde{\alpha}_{j},\tilde{%
\beta}_{j}\right) $, for a sequence of $n$ inverse actions we have $%
\boldsymbol{U}_{n}^{-1}\left( \vec{\alpha},\vec{\beta}\right)
=\prod_{j=n}^{1}\boldsymbol{U}\left( \tilde{\alpha}_{j},\tilde{\beta}%
_{j}\right) $, however as $\tilde{\alpha}_{j}^{2}+\tilde{\beta}%
_{j}^{2}=\left( \alpha _{j}^{2}-\beta _{j}^{2}\right) ^{-2}\neq 1$,
therefore normalization is not possible.

\subsection{Invertibility and complex coefficients}

In order to establish the invertibility of $\boldsymbol{U}\left( \alpha
,\beta \right) $, the domain of $\alpha $ and $\beta $ must be be extended
to the field of complex numbers because the conditions $\left\vert \alpha
\right\vert ^{2}+\left\vert \beta \right\vert ^{2}=1$ and $\alpha ^{2}-\beta
^{2}=1\Longrightarrow \left\vert \beta \right\vert ^{2}+\beta ^{2}=0$ must
be satisfied. This happens for $\alpha $ real and $\beta =-i\left\vert \beta
\right\vert $, a pure imaginary. Since one is left with one free parameter,
a natural parametrization is $\alpha =\cos \xi $ and $\beta =-i\sin \xi $ ($%
\xi $ real), thus $\boldsymbol{U}\left( \alpha ,\beta \right) \equiv 
\boldsymbol{U}\left( \xi \right) =\cos \xi \ \boldsymbol{I}-i\sin \xi \ 
\boldsymbol{X}$ \ is a unitary operator mapping a Cbit or a qubit into a
qubit, $\boldsymbol{U}\left( \xi \right) \left\vert x_{0}\right\rangle =\cos
\xi \ \left\vert x_{0}\right\rangle -i\sin \xi \ \left\vert \bar{x}%
_{0}\right\rangle $. So, the complex nature of $\boldsymbol{U}\left( \xi
\right) $ is due to its invertibility property. A sequence of actions 
\begin{equation}
\boldsymbol{U}_{n}\left( \vec{\xi}\right) =\prod_{j=1}^{n}\left( \cos \xi
_{j}\ \boldsymbol{I}-i\sin \xi _{j}\ \boldsymbol{X}\right) ,  \label{U-1n}
\end{equation}%
on a Cbit $\left\vert x_{0}\right\rangle $ takes it to the qubit $%
\boldsymbol{U}_{n}\left( \vec{\xi}\right) \left\vert x_{0}\right\rangle
=A_{n}\left( \vec{\xi}\right) \left\vert x_{0}\right\rangle +B_{n}\left( 
\vec{\xi}\right) \left\vert \bar{x}_{0}\right\rangle =\left\vert \psi
_{n}\right\rangle $, with coefficients $A_{n}\left( \vec{\xi}\right) =\cos
\left( \sum_{j=1}^{n}\xi _{j}\right) $ and $B_{n}\left( \vec{\xi}\right)
=-i\sin \left( \sum_{j=1}^{n}\xi _{j}\right) $. The parameters $\xi _{j}$
are undetermined and their sum is $\phi _{n}=\sum_{j=1}^{n}\xi _{j}$, so one
can write (\ref{U-1n}) in the compact form%
\begin{equation}
\boldsymbol{U}_{n}\left( \vec{\xi}\right) \Longrightarrow \boldsymbol{U}%
\left( \phi _{n}\right) =\exp \left[ -i\phi _{n}~\boldsymbol{X}\right] ,
\label{Uexp}
\end{equation}%
where $\phi _{n}$ is interpreted as a register parameter, it sets the
ordering of the actions. Due to the indetermination of the parameters $\xi
_{j}$ nothing can be said about the intervals between consecutive actions,
see Figure (\ref{sequen}-a),\FRAME{ftbpFU}{2.8784in}{1.6571in}{0pt}{\Qcb{%
{\protect\small (a) Undetermined intervals between sequences of actions. (b)
Uniformization of the intervals.}}}{\Qlb{sequen}}{sequencias 2.jpg}{\special%
{language "Scientific Word";type "GRAPHIC";display "USEDEF";valid_file
"F";width 2.8784in;height 1.6571in;depth 0pt;original-width
9.5998in;original-height 7.1996in;cropleft "0";croptop "1";cropright
"1";cropbottom "0";filename 'sequencias 2.jpg';file-properties "XNPEU";}}
the vertical bars stand for each action, they can be distributed at will,
although obeying an ordered sequence. Imposing the composition law (\ref%
{trans1})\ one has $\boldsymbol{U}\left( \phi _{n}\right) \boldsymbol{U}%
\left( \phi _{m}\right) =\boldsymbol{U}\left( \phi _{n+m}\right) $, and the
form (\ref{Uexp}) implies $\phi _{n}+\phi _{m}=\phi _{n+m}$; as so,
necessarily and uniquely $\phi _{n}$ must be linear in $n$, $\phi _{n}=n\bar{%
\xi}$ with $\bar{\xi}$ some parameter. Thus $\boldsymbol{U}\left( \phi
_{n}\right) $ becomes $\boldsymbol{U}\left( n\bar{\xi}\right) =\exp \left[
-in\bar{\xi}~\boldsymbol{X}\right] $, which stands for a sequence of
actions, or an evolution. The previously undetermined intervals between
actions become equally spaced, see Figure (\ref{sequen}-b), characterizing
the \emph{uniformization} of their distribution. In order to turn the
distribution dense I shall look for a differential equation for $\boldsymbol{%
U}\left( n\bar{\xi}\right) $ by taking first the difference between two
consecutive values of $n$ and then dividing by $\bar{\xi}$, 
\begin{equation*}
\frac{\boldsymbol{U}\left( \left( n+1\right) \bar{\xi}\right) -\boldsymbol{U}%
\left( n\bar{\xi}\right) }{\bar{\xi}}=\left( \frac{e^{-i\bar{\xi}\boldsymbol{%
X}}-1}{\bar{\xi}}\right) \exp \left[ -in\bar{\xi}~\boldsymbol{X}\right] .
\end{equation*}%
The limit to a continuous parameter is obtained for $n\gg 1$ and $\bar{\xi}%
\ll 1$, keeping however the product $n\bar{\xi}=\tau $ finite. A linear
differential equation results, $id\boldsymbol{U}\left( \tau \right) /d\tau =%
\boldsymbol{XU}\left( \tau \right) $, and $\boldsymbol{U}\left( \tau \right)
=e^{-i\tau \boldsymbol{X}}$, where $\tau $ is the continuous ordering
parameter of the actions, or a local\emph{\ time}\textit{\ }in arbitrary
units, that should be set according to the clock to be used. Writing $%
\left\vert x_{\tau }\right\rangle =\boldsymbol{U}\left( \tau \right)
\left\vert x_{0}\right\rangle $ the evolution equation $id\left\vert x_{\tau
}\right\rangle /d\tau =\boldsymbol{X}\left\vert x_{\tau }\right\rangle $
says how a qubit evolves due to the action of $\boldsymbol{X}$, which is the
generator of the changes.

Defining a more general generator, $\boldsymbol{G}=\mu \boldsymbol{I}+\nu 
\boldsymbol{X}$, $\mu $ and $\nu $ being two real parameters, the evolution
equation writes 
\begin{equation}
i\frac{d\left\vert \psi _{\tau }\right\rangle }{d\tau }=\boldsymbol{G}%
\left\vert \psi _{\tau }\right\rangle ,  \label{eqsch2}
\end{equation}%
with $\boldsymbol{U}\left( \tau \right) =e^{-i\tau \mu \boldsymbol{I}%
}e^{-i\tau \nu \boldsymbol{X}}$ for the evolution operator. Differently from
the factor $e^{-i\tau \nu \boldsymbol{X}}$ that do really affect the
evolution of a qubit, the phase factor $e^{-i\tau \mu \boldsymbol{I}}$ is
apparently no significant because, besides a global phase factor, it does
not entail any change when acting on Cbit or qubit. The eigenvalues and
eigenstates of $\boldsymbol{G}$ are respectively $G_{\pm 1}=\mu \pm \nu $, $%
\left\vert x_{\pm 1}\right\rangle =\left( \left\vert 0\right\rangle \pm
\left\vert 1\right\rangle \right) /\sqrt{2}$. A general solution to Eq. (\ref%
{eqsch2}) is $\left\vert \psi _{\tau }\right\rangle =\sum_{\sigma =\pm
1}e^{-iG_{\sigma }\tau }c_{\sigma }\left\vert x_{\sigma }\right\rangle $
where $G_{\sigma }=\mu +\sigma \nu $. Now conjecturing about the qubit
carrier, I assume it a massive particle \cite{LANDAUER91} and the parameter $%
\mu $ is chosen to represent its energy; thus the change $\boldsymbol{%
X\rightarrow }$ $\boldsymbol{G}$ is important because it allows the
introduction of that particle property. $\boldsymbol{G}$ can be identified
as a hamiltonian, and for an arbitrary initial condition the mean value is $%
\left\langle \psi _{\tau }\right\vert \boldsymbol{G}\left\vert \psi _{\tau
}\right\rangle =\mu +\nu \left( \left\vert c_{+1}\right\vert ^{2}-\left\vert
c_{-1}\right\vert ^{2}\right) $; while $\mu $ is the particle kinetic
energy, the second term is the qubit energy that exists only when it is
coupled to some field ($\nu \neq 0$). {\LARGE \ }

\section{The qubit carrier and the Pauli-Schr\"{o}dinger equation}

The spatial localization of the carrier must be introduced into Eq. (\ref%
{eqsch2}), thus for a qubit state $\left\vert \psi _{0}\right\rangle
=a_{0}\left\vert x_{0}\right\rangle +b_{0}\left\vert \bar{x}%
_{0}\right\rangle $ the parameters $a_{0},b_{0}$ should depend on the
position $q$, namely, $\left\vert \psi _{0}\left( q\right) \right\rangle
=a_{0}\left( q\right) \left\vert x_{0}\right\rangle +b_{0}\left( q\right)
\left\vert \bar{x}_{0}\right\rangle $ becomes the state of the whole system,
with normalization $\int dq\ \left\vert a_{0}\left( q\right) \right\vert
^{2}+\int dq\ \left\vert b_{0}\left( q\right) \right\vert ^{2}=1$. The qubit
state is correlated to the particle position that influences its probability
outcomes $\left\vert a_{0}\left( q\right) \right\vert ^{2}$ and $\
\left\vert b_{0}\left( q\right) \right\vert ^{2}$. Coordinate dependence
should also be present in the generator, so $\boldsymbol{G}\left( q\right)
=\mu \left( q\right) \boldsymbol{I}+\nu \boldsymbol{X}$ and the parameter $%
\nu $ is assumed $q$-independent because interaction between both degrees of
freedom is not considered. The evolved state is $\boldsymbol{U}\left( \tau
\right) \left\vert \psi _{0}\left( q\right) \right\rangle =\left\vert \psi
\left( q,\tau \right) \right\rangle =a_{\tau }\left( q\right) \left\vert
x_{0}\right\rangle +b_{\tau }\left( q\right) \left\vert \bar{x}%
_{0}\right\rangle $, with amplitudes 
\begin{eqnarray}
a_{\tau }\left( q\right) &=&e^{-i\tau \mu \left( q\right) }\left(
a_{0}\left( q\right) \cos \nu \tau -ib_{0}\left( q\right) \sin \nu \tau
\right) \   \label{atau} \\
b_{\tau }\left( q\right) &=&e^{-i\tau \mu \left( q\right) }\left(
-ia_{0}\left( q\right) \sin \nu \tau +b_{0}\left( q\right) \cos \nu \tau
\right) .  \label{btau}
\end{eqnarray}%
with $a_{0}\left( q\right) $ and $b_{0}\left( q\right) $ as initial values.
So, the qubit was merged with the spatial motion of its carrier within a
single equation, meaning that the joint evolution -- the qubit sequence of
actions as well as the change in the spatial configuration of the carrier --
is measured by a single clock. To determine the parameters $a_{0}\left(
q\right) $ and $b_{0}\left( q\right) $ they should obey some differential
equation for the variable $q$, then $\mu \left( q\right) $ must depend also
on $\partial /\partial q$ and/or its powers. However, instead of trying to
guess the functional form, it is better to take advantage of the available
information from hamiltonian mechanics, so I define $\mu $ as the kinetic
energy of a non-relativistic particle, $\mu \Longrightarrow T\left( p\right) 
$ $=p^{2}/2m$, where $p$ is the linear momentum in some reference frame. Eq.
(\ref{eqsch2}) becomes $i\kappa _{0}d\left\vert \tilde{\psi}\left( p,\tau
\right) \right\rangle /d\tau =\left[ T\left( p\right) \boldsymbol{I}%
+\varepsilon _{0}^{\prime }\nu \boldsymbol{X}\right] \left\vert \tilde{\psi}%
\left( p,\tau \right) \right\rangle $. Since $T\left( p\right) $ has units
of energy, the second term in brackets should also have the same units. So
the constants $\kappa _{0}$ and $\varepsilon _{0}^{\prime }$, have both
units of energy. One can also choose some unit to measure the dimensionless
time $\tau $, $\tau =t/t_{0}$, so the dynamical equation becomes%
\begin{equation}
ih_{0}\frac{d\left\vert \tilde{\psi}\left( p,t\right) \right\rangle }{dt}=%
\left[ T\left( p\right) \boldsymbol{I}+\varepsilon _{0}\boldsymbol{X}\right]
\left\vert \tilde{\psi}\left( p,t\right) \right\rangle ,  \label{schreq2}
\end{equation}%
where $h_{0}=\kappa _{0}t_{0}$, $\varepsilon _{0}=\varepsilon _{0}^{\prime
}t_{0}\nu $. Note that the constant $h_{0}$ has units of energy $\times $
time and $\varepsilon _{0}$ has units of energy. An arbitrary initial
condition assumes that the particle momentum and the qubit state are
correlated and the probability amplitude 
\begin{equation}
\left\vert \tilde{\psi}_{0}\left( p,0\right) \right\rangle =\left\vert 
\tilde{\psi}_{0}\left( p\right) \right\rangle =\tilde{a}_{0}\left( p\right)
\left\vert x_{0}\right\rangle +\tilde{b}_{0}\left( p\right) \left\vert \bar{x%
}_{0}\right\rangle ,  \label{supmom}
\end{equation}%
depends on the particle momentum and it contains all the available
information. In momentum space the evolution operator is $\boldsymbol{U}%
\left( t\right) =\exp \left[ -it\left( T\left( p\right) \boldsymbol{I}%
+\varepsilon _{0}\nu \boldsymbol{X}\right) /h_{0}\right] $ and the solution
to Eq. (\ref{schreq2}) is%
\begin{equation*}
\left\vert \tilde{\psi}\left( p,t\right) \right\rangle =e^{-itT\left(
p\right) /h_{0}}\left[ \tilde{a}_{t}\left( p\right) \left\vert
x_{0}\right\rangle +\tilde{b}_{t}\left( p\right) \left\vert \bar{x}%
_{0}\right\rangle \right]
\end{equation*}%
with $\tilde{a}_{t}\left( p\right) =\cos \left( \varepsilon _{0}\nu
t/h_{0}\right) \tilde{a}_{0}\left( p\right) -i\sin \left( \varepsilon
_{0}\nu t/h_{0}\right) \tilde{b}_{0}\left( p\right) $ and $\tilde{b}%
_{t}\left( p\right) =\cos \left( \varepsilon _{0}\nu t/h_{0}\right) \tilde{b}%
_{0}\left( p\right) -i\sin \left( \varepsilon _{0}\nu t/h_{0}\right) \tilde{a%
}_{0}\left( p\right) $, and the particle mean energy is $\left\langle \tilde{%
\psi}\left( p,t\right) \right\vert \boldsymbol{H}\left( p\right) \left\vert 
\tilde{\psi}\left( p,t\right) \right\rangle =T\left( p\right) +2\varepsilon
_{0}{Re}\left( \tilde{a}_{0}^{\ast }\left( p\right) \tilde{b}_{0}\left(
p\right) \right) $. Since coordinate and momentum are conjugated variables
the statevector in coordinate representation is 
\begin{equation*}
\left\vert \psi \left( q,t\right) \right\rangle =\psi _{x_{0}}\left(
q,t\right) \left\vert x_{0}\right\rangle +\psi _{\bar{x}_{0}}\left(
q,t\right) \left\vert \bar{x}_{0}\right\rangle ,
\end{equation*}%
and $\psi _{{x_{0}}}\left( q,t\right) $, $\psi _{\bar{x}_{0}}\left(
q,t\right) $ are the amplitudes associated to the Cbits $\left\vert
x_{0}\right\rangle $, $\left\vert \bar{x}_{0}\right\rangle $;$\ $they can be
written as Fourier transforms 
\begin{equation}
\psi _{\binom{x_{0}}{\bar{x}_{0}}}\left( q,t\right) =\int \frac{dp}{2\pi }%
e^{ipq/h_{1}}e^{-itT\left( p\right) /h_{0}}\binom{\tilde{a}_{t}\left(
p\right) }{\tilde{b}_{t}\left( p\right) }.  \label{schreq22}
\end{equation}%
The constant $h_{1}$ is introduced to set the correct dimensionality, it has
the same units as $h_{0}$, nonetheless nothing can be said about being the
same constant, unless confirmed by experiment. In Eq. (\ref{schreq22}) 
\begin{equation}
\binom{\tilde{a}_{t}\left( p\right) }{\tilde{b}_{t}\left( p\right) }=\int
dq^{\prime }e^{-ipq^{\prime }/h_{1}}\binom{a_{t}\left( q^{\prime }\right) }{%
b_{t}\left( q^{\prime }\right) },  \label{psi02}
\end{equation}%
with $a_{t}\left( q^{\prime }\right) =\cos \left( \varepsilon _{0}\nu
t/h_{0}\right) a_{0}\left( q^{\prime }\right) -i\sin \left( \varepsilon
_{0}\nu t/h_{0}\right) b_{0}\left( q^{\prime }\right) $ and $b_{t}\left(
q^{\prime }\right) =\cos \left( \varepsilon _{0}\nu t/h_{0}\right)
b_{0}\left( q^{\prime }\right) -i\sin \left( \varepsilon _{0}\nu
t/h_{0}\right) a_{0}\left( q^{\prime }\right) $. So even not existing a
direct interaction between the qubit and its carrier, the probability for
measuring the qubit in Cbit $\left\vert x_{0}\right\rangle $, or $\left\vert 
\bar{x}_{0}\right\rangle $, becomes affected by its position.

Using Eqs. (\ref{schreq22}) and (\ref{psi02}) and manipulating Eq. (\ref%
{schreq2}) it is not hard to verify that one can substitute the c-number $p$
by the derivative $-ih_{1}\partial /\partial q$, and we can rewrite that
equation as 
\begin{equation}
ih_{0}\frac{\partial \left\vert \psi \left( q,t\right) \right\rangle }{%
\partial t}=\left[ \frac{1}{2m}\left( -ih_{1}\frac{\partial }{\partial q}%
\right) ^{2}\boldsymbol{I}+\varepsilon _{0}\nu \boldsymbol{X}\right]
\left\vert \psi \left( q,t\right) \right\rangle .  \label{schreq3}
\end{equation}%
The terms in brackets stand for the particle and qubit hamiltonian in
coordinate and matrix representation, so the parameter $\mu $ becomes
determined. In the presence of an energy conserving potential $V\left(
q\right) $ the PSE takes its familiar form, with hamiltonian $\boldsymbol{H}%
=H_{0}\boldsymbol{I}+\varepsilon _{0}\nu \boldsymbol{X}$ and $H_{0}=\left[
(-ih_{1}/(2m))({\partial }^{2}/{\partial ^{2}q})+V\left( q\right) \right] $.
The particle described by Eq. (\ref{schreq3}) has now blurred classical
properties (it looses the sharp trajectory it has in phase space), its best
representation is a wavefunction and the appearance of quantum properties
are due to the qubit it is carrying. Any further generalization is trivial
and immediate: (1) from 1-D to 3-D in spacial coordinates, $\partial
/\partial q\rightarrow \nabla $ and (2) since any $2\times 2$ matrix can be
expanded in the basis formed by the unit matrix $\boldsymbol{I}$ and Pauli
matrices ($\sigma _{x}$, $\sigma _{y}$, and $\sigma _{z}$), then $\nu 
\boldsymbol{X}\longrightarrow \vec{\nu}\cdot \boldsymbol{\vec{\sigma}}$.

\section{Dirac equation: two qubits of information}

Few words about Dirac equation $i\hbar \partial \left\vert \Psi _{D}\left(
t\right) \right\rangle /\partial t=\boldsymbol{H}_{D}\left\vert \Psi
_{D}\left( t\right) \right\rangle $, its hamiltonian is $\boldsymbol{H}_{D}=c%
\boldsymbol{\vec{\alpha}}\cdot \vec{p}+mc^{2}\boldsymbol{\beta }$ and the
four dimension-4 matrices $\boldsymbol{\vec{\alpha}}$, $\boldsymbol{\beta }$
\ satisfy the relations $\boldsymbol{\alpha }_{k}\boldsymbol{\alpha }_{l}+%
\boldsymbol{\alpha }_{l}\boldsymbol{\alpha }_{k}=2\boldsymbol{I}\delta _{kl}$%
, $\boldsymbol{\vec{\alpha}\beta }+\boldsymbol{\beta \vec{\alpha}=0}$ and $%
\boldsymbol{\beta }^{2}=0$. These matrices can be expressed as tensor
products of dimension-2 matrices, each one acting on its own qubit, $%
\boldsymbol{\vec{\alpha}\ =\ }\boldsymbol{\boldsymbol{X}_{1}\otimes \left( 
\boldsymbol{X}_{2},i\boldsymbol{Y}_{2},\boldsymbol{Z}_{2}\right) =X}%
_{1}\otimes \boldsymbol{\vec{\sigma}}_{2}$, so $c\boldsymbol{\vec{\alpha}}%
\cdot \vec{p}=\boldsymbol{X}_{1}\otimes \left( c\vec{p}\cdot \boldsymbol{%
\vec{\sigma}}_{2}\right) $ and $\boldsymbol{\beta =Z}_{1}\otimes \boldsymbol{%
I}_{2}.$ Thus, Dirac's hamiltonian can be written as tensor products acting
on independent D-2 Hilbert subspaces $\boldsymbol{H}_{D}=\boldsymbol{Z}%
_{1}\otimes \left( mc^{2}\boldsymbol{I}_{2}\right) +\boldsymbol{X}%
_{1}\otimes \left( c\vec{p}\cdot \boldsymbol{\vec{\sigma}}_{2}\right) $.
Squaring $\boldsymbol{H}_{D}$ one gets the relativistic energy $\left( 
\boldsymbol{H}_{D}\right) ^{2}=E_{p}^{2}\ \boldsymbol{I}_{1}\otimes 
\boldsymbol{I}_{2}$, where $E_{p}^{2}=m^{2}c^{4}+c^{2}\vec{p}^{\ 2}$. The
time-dependent equation reduces into direct products of $2\times 2$ matrices 
\begin{eqnarray*}
&&\left[ \boldsymbol{I}_{1}\otimes \left( i\hbar \frac{\partial }{\partial t}%
\boldsymbol{I}_{2}\right) -\boldsymbol{Z}_{1}\otimes \left( mc^{2}%
\boldsymbol{I}_{2}\right) -\boldsymbol{X}_{1}\otimes \left( c\vec{p}\cdot 
\boldsymbol{\vec{\sigma}}_{2}\right) \right] \\
&\times &\left\vert \Psi _{D}\left( t\right) \right\rangle =0,
\end{eqnarray*}%
which is invariant under Lorentz transformation. The solutions are 
\begin{eqnarray}
\left\vert \Psi ^{\lambda }\left( \vec{p},t\right) \right\rangle
&=&N_{\lambda }e^{-i\lambda tE_{p}}\left[ \left\vert 1\right\rangle
_{1}\left\vert \varphi \left( \vec{p}\right) \right\rangle _{2}\right. 
\notag \\
&+&\left. \left\vert 0\right\rangle _{1}\frac{c\vec{p}\cdot \boldsymbol{\vec{%
\sigma}}_{2}}{mc^{2}+\lambda E_{p}}\left\vert \varphi \left( \vec{p}\right)
\right\rangle _{2}\right] ,  \label{fundirac}
\end{eqnarray}%
where 
\begin{equation*}
\left\vert \varphi \left( \vec{p}\right) \right\rangle _{2}=\left( 
\begin{array}{c}
\varphi _{+}\left( \vec{p}\right) \\ 
\varphi _{-}\left( \vec{p}\right)%
\end{array}%
\right) _{2}
\end{equation*}%
with $\lambda =\pm 1$ standing for positive and negative energy solutions
and $N_{\lambda }$ is a normalization constant. The qubit 2 in Eq. (\ref%
{fundirac}) represents the particle state whereas the Cbit 1 is apparently
ancillary, it works as a selector: the projector $\left( \left\vert
1\right\rangle \left\langle 1\right\vert \right) _{1}$ selects the
nonrelativistic component $\left\vert \varphi \left( \vec{p}\right)
\right\rangle _{2}$ while $\left( \left\vert 0\right\rangle \left\langle
0\right\vert \right) _{1}$ projects the relativistic complement. Also
interesting is that all the $\gamma _{\mu }$ matrices have the structure of
the direct product of two-qubit operators $\gamma ^{0}=\boldsymbol{Z}%
_{1}\otimes \boldsymbol{I}_{2}$, $\gamma ^{1}=i\boldsymbol{Y}_{1}\otimes 
\boldsymbol{X}_{2}$, $\gamma ^{2}=-\boldsymbol{Y}_{1}\otimes \boldsymbol{Y}%
_{2}$, $\gamma ^{3}=i\boldsymbol{Y}_{1}\otimes \boldsymbol{Z}_{2}$. $\quad $

\section{Concluding remarks}

As long as the qubit is not probed (for a spin, there is no external
magnetic field), $\nu =0$, Eq. (\ref{schreq3}) reduces to two uncoupled Schr%
\"{o}dinger equations 
\begin{equation}
ih_{0}\frac{\partial }{\partial t}\left( 
\begin{array}{c}
\psi _{x_{0}}\left( q,t\right) \\ 
\psi _{\bar{x}_{0}}\left( q,t\right)%
\end{array}%
\right) =H_{0}\left( 
\begin{array}{c}
\psi _{x_{0}}\left( q,t\right) \\ 
\psi _{\bar{x}_{0}}\left( q,t\right)%
\end{array}%
\right) ,  \label{schreq5}
\end{equation}%
for two wavefunctions $\psi _{x_{0}}\left( q,t\right) $, $\psi _{\bar{x}%
_{0}}\left( q,t\right) $ in one qubit/spin eigenstate. The two equations are
redundant thus the relevant information resides in one of them only, going
to the usual spinless SE. Although classical physics was crucial to arrive
at Eq. (\ref{schreq5}), by setting $\nu =0$ the particle motion is not ruled
(back) by classical physics (Hamilton equations) but by the usual SE. So why
do quantum properties of the particle still persist even when the
correlation between a qubit/spin and its carrier is broken? The answer is
that even not being activated the qubit/spin is still carried by the
particle, the internal degree of freedom and the particle make one single
object, although not entangled. One is left with an equation (SE) that does
not keep any clue about the presence of a qubit/spin, nonetheless it is
still there although not manifestly evident. That%
%TCIMACRO{\U{b4}}%
%BeginExpansion
\'{}%
%EndExpansion
s why the SE can be used without any mention to spin if not needed;
otherwise, this internal degree of freedom must be appended in order to
explain observed phenomena. In conclusion, because it is carrying one qubit
of information, to the observer the particle shifts its behavior from the
classical picture, it acquires wave properties with a probabilistic
character where the uncertainty relations represent one facet.

\begin{acknowledgments}
Acknowledgements
\end{acknowledgments}

I wish to express thanks for the support from FAPESP and CNPq, Brazilian
agencies.

\textbf{Figure captions}

Fig. 1. (a){\small \ }Undetermined intervals between sequences of actions.
(b) Uniformization of the intervals.

\end{document}